# Towards Achieving Trust Through Transparency and Ethics

David Kwan and Luiz Marcio Cysneiros
School of Information Technology
York University
Toronto, Canada
dkwan33@yorku.ca, cysneiro@yorku.ca

Julio Cesar Sampaio do Prado Leite
Dep. De Informática PUC
Pontificia Universidade Catolica
Rio de Janeiro, Brazil
julio@inf.puc-rio.br

*Abstract*— The ubiquitous presence of software in the products we use, together with Artificial Intelligence in these products, has led to an increasing need for consumer trust. Consumers often lose faith in products, and the lack of Trust propagates to the companies behind them. This is even more so in mission-critical systems such as autonomous vehicles and clinical support systems. This paper follows grounded theory principles to elicit knowledge related to Trust, Ethics, and Transparency. We approach these qualities as Non-Functional Requirements (NFRs), aiming to build catalogs to subsidize the construction of Socially Responsible Software. The corpus we have used was built on a selected collection of literature on Corporate Social Responsibility, with an emphasis on Business Ethics. Our challenge is how to encode the social perspective knowledge, mainly through the view of Corporate Social Responsibility, on how organizations or institutions achieve trustworthiness. Since our ground perspective is that of NFRs, results are presented by a catalogue of Trust as a Non-Functional Requirement, represented as a Softgoal Interdependency Graph (SIG). The SIG language helps software engineers in understanding alternatives they have to improve Trust in software products.

I. INTRODUCTION

Software today is omnipresent in daily life. Transparency is becoming more and more important to consumers, which extends to the final product and the developmental process itself [1]. Consumers need to trust software and software developers [2], but this Trust is frequently violated, leading to loss of Trust. This can be seen clearly in the recent decline of consumer trust in Facebook due to breaches on its users' personal information. This led Facebook to respond with apologies through major UK and US media [3]. Being transparent in their late apology[1], Facebook aimed to retain consumers. On another front, AI advances, including the current high-profile issue of autonomous vehicles, lead to complex ethical dilemmas that impair user trust.

Furthermore, this lack of clarity extends to the boundaries of legal responsibility and the responsibility of parties involved in legal disputes. In the event of collisions leading to loss of property, or worse, loss of life, the party responsible for reparations is unclear [4]. The Mercedes-Benz automotive corporation has announced that its self-driving vehicles will be designed to prioritize occupant safety over pedestrian safety [5][2]. At a cursory glance, this appears to be the correct approach to appease consumer concerns regarding the driver's safety; after all, who would buy a car that would willingly endanger the driver? The ethics and legality behind such a decision are not so simple. This is an example of a jealous algorithmic approach, and Mercedes-Benz's decision to use such an approach may be ethically dubious and result in legal problems [6]. If companies want their products to be successful, they must consider the effects of their handling of Transparency and ethical decisions on the product's trustworthiness. This is especially true for mission-critical systems such as autonomous vehicles or clinical support systems in the healthcare domain [7], [8].

We agree with the idea of Socially Responsible Software (SRS) [9], which holds that the software development process needs to address Trust in software-embedded products. Within SRS, Trust in software, from the consumer viewpoint, was inspired by the metaphor of Corporate Social Responsibility (CSR) [9]. CSR promotes loyalty, repeat business, and purchase intention [10]. CSR also promotes intrinsic Trust and positively impacts market value and companies' profits perceived to be committed to CSR [11]. Cysneiros and Leite developed an initial Softgoal Interdependency Graph (SIG) [12] illustrating the key non-functional requirements (NFRs) identified for creating SRS [9]. Their SIG introduces how Ethics and Transparency are two NFRs vital to the fostering of Trust.

We extended this SIG knowledge to further represent how we can foster Trust and create trustworthy software. Much of the literature in this domain is focused on specific circumstances such as trust repair in cases of lost Trust, social responsibility and its relation between companies and stakeholders, or in specific fields such as autonomous vehicles.

This work aims to build upon the literature and develop a comprehensive view, providing several possible solutions software engineers can use to develop software that can be trusted and, therefore, motivate its use by society and eventually help companies be trusted, attracting new maintaining the current clients. Moreover, this work put together emerging concepts in requirements engineering such as Transparency and Ethics with Trust that, despite the existing work, is still a very open area mostly when used in the context of business and social concepts as it is done here. It also brings many possible realistic approaches that requirements engineers can use to elicit and model solutions that can lead to Socially Responsible Software development.

---

[1] The breach occurred in 2014 and the apology was published in March 2018.

[2] In 2012, the auto industry produced the first car with pedestrian airbag, the Volvo V40.

II. RESEARCH APPROACH

This work aims to provide a body of knowledge to facilitate the understanding and delivery of trustworthy software. The social factors that affect Trust related to the corporate world and its impact on companies and institutions are relatively studied. In contrast, Trust in the software domain is narrower and not focused on attracting new clients while maintaining existing ones. Thus, we primarily sample from the business domain's knowledge of CSR to understand the factors that influence trustworthiness, while also focusing on the NFRs of Transparency and Ethics related to the field of information technology. Transparency and Ethics were chosen because they are a pillar to achieve CSR.

To address the challenge of encoding the social perspective knowledge of Trust, we relied on a social science elicitation method (Grounded Theory (GT)), and a requirements engineering modeling framework, the NFR framework [12]. Our work used a previous Trust NFR catalog [9] and created a translation protocol from the concepts elicited using GT to evolve the Trust catalog. An NFR catalogue is a collection of SIGs that showcase the relationships between multiple softgoals and their operationalization to satisfice one particular NFR. SIGs, being conceptual models of the NFR framework, are well suited to abstracting, comprehending, and communicating knowledge [12], [13]. As Aydemir and Dalpiaz illustrate in their roadmap for Ethics-aware software engineering, visual notations communicate information better than text [13].

Grounded theory (GT) is a qualitative research methodology that originated in the social sciences [14]. It has been used in requirements engineering due to its suitability for discovering patterns, thus helping the elicitation of concepts [15], [16]. Furthermore, this approach provides a structured and traceable method for generating theory directly from unstructured qualitative data sources through a continuous interplay between data collection and data analysis [14], [17].

The use of GT in requirements engineering has been combined with different elicitation techniques such as interviews and observations [16], as well as questionnaires (surveys) [18]. Independent of the elicitation technique, GT uses data collected as transcripts (a corpus), that is text. In our case, we used the texts of the CSR related literature that we have selected as our corpus. We elicited NFRs concepts from the corpus following Wolfswinkel et al.'s approach to using grounded theory as a literature review method [17], which is directly based on the principles of grounded data analysis outlined by Strauss and Corbin [14]. However, we did not limit our search to CSR literature. We searched computer science literature also departing from terms used in the CSR literature.

The corpus was built using keyword search on Google Scholar and IEEE Xplore. We defined the search string as such: ("trust" OR "trustworthiness" OR "transparency" OR "ethics" OR "ethical") AND ("design" OR "software" OR "information technology" OR "legal" OR "legality" OR "regulation" OR "corporate social responsibility" OR "corporate social performance"). For the inclusion criteria, we included papers from peer-reviewed sources and written in English. For the exclusion criteria, we excluded duplicate results and papers outside the scope (by title and main topics) of this research. A total of 103 publications were acquired, and after further reading of the abstracts, 36 articles were selected for the coding stages of grounded theory-based literature review. Papers were deemed to be within scope if the content discussed relationships or phenomenon pertaining to Trust and Transparency, Trust and Ethics, or Trust, Transparency and Ethics together. Two of the authors collaborated close to selecting these papers. Of the 36 articles selected, all of them are used as a reference in this paper. More specifically, these papers are: [2], [3], [6]–[8], [10], [11], [19]–[47]. These papers are also explicitly linked to the SIGs' subgoals and operationalizations, as is explained in Section 3.

Wolfswinkel et al.'s approach is an iterative method with three primary steps: open coding, axial coding, and selective coding [17]. Each coding step involves creating salient categories, sub-categories, or relationships from analysis of the literature. The first stage, open coding, involves analyzing each article for salient concepts and identifying and naming conceptual categories [14], [17]. As these concepts were made aware to us, we recorded each concept's association with its corresponding article in a matrix. A glimpse of the open coding matrix is shown in Table I, as an example. Throughout the coding process, concepts were continually added, merged, or separated based on multiple evaluation cycles. As new concepts were identified, they were compared to the set of existing concepts, often resulting in revision or consolidation of existing concepts. Concepts that were found to be intricately related and merged were grouped into a single concept. Whenever a concept was re-evaluated, all other concepts were re-evaluated to determine whether the initial re-evaluation warranted any further changes.

TABLE I. OPEN CODING MATRIX (PARTIAL)

|  | Trust | Ethics | Transparency in Operations | CSR |
|---|---|---|---|---|
| **Bews 2002** | x | x |  |  |
| **Bussone 2015** | x |  |  |  |
| **Pivato 2007** | x |  | x | x |

The second stage of our grounded theory-based analysis, axial coding, involved creating new concepts (Wolfswinkel et al.'s sub-categories) and relating them to each other as well as the categories from open coding [17]. The purpose of axial coding is to further understand the phenomenon under investigation by making connections between concepts (seen as softgoals). These new concepts included any type of solution or activity that could be utilized to satisfice one or more concepts or even the primary concept (softgoal) of Trust itself. These *solutions* are also referred to as *operationalizations*. Rather than use the term "satisfy", since an NFR can rarely be fully satisfied, we use the term "satisfice" to indicate that the NFR is satisfied to a reasonable degree [1]. This relationship could include causal conditions, phenomena, strategies, and context.

The third stage of our analysis, selective coding, involved integrating the concepts from open coding and the concepts and operationalizations from axial coding with our central goal of Trust. The purpose of selective coding is to relate categories and sub-categories to a single central concept [14]. The relationships between each pair: concept and operationalization, concept and concept, operationalization and operationalization were determined throughout each evaluation cycle. Each relationship pairing was recorded with a contribution type (positive, negative) and level of severity (break, hurt, help, make).

TABLE II. AXIAL AND SELECTIVE CODING MATRIX (PARTIAL)

| Operationalization or Subgoal | Contribution Type | Severity | Target | Source |
|---|---|---|---|---|
| Transparency in Operations | Positive | Help | Transparency | Bews 2002<br>Bussone 2015<br>*[sources truncated]* |
| Openness | Positive | Help | Ethical Principles | Bews 2002<br>Schoorman 2007 |
| Integrity | Positive | Help | Ethical Principles | Bews 2002<br>Schoorman 2007 |
| Overreliance | Negative | Hurt | Trust | Bussone 2015 |

As outlined by Wolfswinkel et al., the coding iterations are not rigid steps [17]. We traversed back and forth between the articles, concepts, and operationalizations to continuously refine and elaborate upon them. The coding process stopped when we were satisfied that any new re-readings did not provide further insight into any concept, nor did it provide further insight into new salient concepts [17]. Traceability, a vital component of grounded theory-based literature review, was maintained throughout the entire coding process. Traceability links between concepts or operationalizations with their source article are explicitly shown in our results. Since axial and selective coding steps are so heavily intertwined, they were marked in one matrix. A partial example of the final coding matrix is shown in Table II. The full matrices described in Table I and Table II can be found at doi:10.5281/zenodo.5076626.

Using the existing Trust catalogue from [9] as a basis, the matrices created from our grounded theory analysis were translated to the SIG language [12], upstanding the Trust catalogue from [9]. One researcher performed the translation step, while two other researchers reviewed the results. Categories elicited from open coding mapped directly to primary softgoals (such as Trust or Ethics) and provided clear lines of division to separate our catalogue into SIGs. Concepts elicited from axial coding mapped to either operationalizations or SIGs, depending on the context within the literature. If the literature defines the concept as a course of action, then it mapped to an *operationalization*; otherwise it mapped to a *softgoal*. Connections between concepts elicited from either axial coding or selective coding mapped to the relationships (arrows) between elements of the SIGs.

Many operationalizations could ambiguously fall into multiple categories due to the complex intertwined nature of the relationships studied. Hence, a judgment call was made based on the primary goal or focus of the operationalization as to which category it belonged to. When necessary, these operationalizations were included in multiple SIGs, where the operationalization from one category appears in another SIG context. This is done to improve the comprehensibility of the SIGs. Although this has some redundancy, this facilitates readability and helps the reader keep track of the relationships between the various operationalizations and subgoals.

During the stage of selective coding, we opted to primarily use contribution decompositions to develop relationships to ease comprehension in the SIGs. Contribution links were marked in the matrix between operationalizations and softgoals to any other operationalizations or softgoals. These contributions can be either positive or negative, and their severity can be one of three: help/hurt (+/- arrows), make/break (++/-- arrows), and some +/- (S+/S- arrows) for positive and negative relationships, respectively. The help/hurt severity indicates that the operationalization or softgoal will have some significant positive or negative contribution to the target, respectively. Contributions were primarily between operationalization and softgoals. However, in some cases, one operationalization may also contribute positively or negatively to other operationalizations. The make/break severity indicates that the operationalization alone is enough to satisfice or prevent the parent softgoal entirely. The some +/- severity indicates a lower-level of certainty and effect than the help/hurt severity. This refers to the certainty present in the literature being analyzed. If the authors of the literature in question seem to be unsure of a certain relationship, then the certainty of the relationship is called into question and thus a different contribution type is used to represent this uncertainty. Table III shows a legend describing SIG notation.

Our catalogue focuses on one high-level NFR of Trust and refines this softgoal into subgoals, which are further refined into identified operationalizations that affect the primary NFR. The catalogues do not intend to provide one single solution. It aims at providing a comprehensive set of solutions. We understand that different domains may require different solutions, and even within the same domain, different projects may target different market shares and hence, require a different solution.

TABLE III. SOFTGOAL INTERDEPENDENCY GRAPH LEGEND

| Element | Description |
|---|---|
| ☁ | Operationalization |
| ☁ | Softgoal (Subgoal) |
| ++→ | Make Contribution |
| +→ | Help Contribution |
| S+→ | Some+ Contribution |
| - -→ | Break Contribution |
| -→ | Hurt Contribution |
| S-→ | Some- Contribution |

## III. RESULTS

### A. High-Level Trust Catalogue

Figure 1 shows a high-level Trust catalogue. This catalogue contains categories of subgoals elicited during the literature review, showing its overall impact on Trust. Each category will be further decomposed and analyzed in the detailed trust catalogues in the sections to follow. Operationalizations extracted from the grounded data analysis present challenges or solutions discussed in the articles that impact Ethics, Transparency, or Trust. These were logically grouped into one of the eight salient categories elicited during open coding. 1) Corporate Social Responsibility, 2) Reputation, 3) Internal Trust, 4) Ethically Motivated Practices, 5) Ethically Motivated Guidelines and Regulation, 6) Ethical Design, 7) Transparency in Operations, and 8) Transparency in Design. Each of these categories can be found on Fig. 1 represented as subgoals. This high-level catalogue displays the relationships between these categories and their impact on the overall goal of Trust. Each of the following detailed catalogues will decompose these categories into their elicited operationalizations. Full-size images of each SIG to facilitate readability can be found at doi:10.5281/zenodo.5076626.

In the following detailed catalogues, operationalizations and some subgoals will have a letter on the SIG that corresponds to a letter at the beginning of the paragraphs of the following text. As such, there is a trace from the text (which summarizes the literature) to the SIG and vice-versa. This facilitates readability by marking the link between the SIG and the elicited facts in the literature. Central softgoals such as Ethics are not explicitly linked since the traceability can be assumed to be all the references used in the corresponding section. Occasionally, an operationalization is not explicitly linked with a letter because it appears across multiple SIGs and the traceability can be found in a different SIG.

### B. Corporate Social Responsibility

(A)The adoption of CSR helps build Trust in an organization, and most, if not, all authors consider it critical to a firm's success. Delgado-Ballester argues that Trust is the most important attribute that any brand can own [45]. Pivato further argues that CSR activities' first result is creating Trust among stakeholders [46]. Furthermore, CSR performance leads to improved stakeholder engagement, which increases mutual Trust and cooperation [19], [20], [47].

(B)Positive social impact results from beneficial CSR practices. The field of business Ethics strongly values social responsibility as a cornerstone of ethical business practice, with much research on how ethical decision making can benefit both social responsibility and profit [21], [22]. Delgado-Ballester states that stakeholders' belief in an organization's positive social impact leads to increased Trust [45].

(C)Corporate Social Performance (CSP) is an extension of Corporate Social Responsibility. CSP's focus is on the actual results achieved instead of the more general idea of responsibility to society that CSR embodies [23]. Pivato found that the perceived CSP of a company by stakeholders, that is, the company's recognition as being socially responsible, results in increased Trust in the company. It is important to note that the stakeholder perception of a company's CSP may not accurately reflect CSP's actual results, nor is it necessarily reflective of a company's commitment to CSR [46].

(D)Pivato also found that efforts that aid CSP monitoring and its effectiveness help CSP's perceived effectiveness [46]. The practices of creating sustainability reports, codes of conduct, labels, proper auditing, and reporting initiatives are all efforts that aid CSP monitoring, and in turn, help perceived CSP [46]. Proper auditing and record-keeping, particularly with codes of ethical guidelines, can help firms navigate governmental guidelines [24].

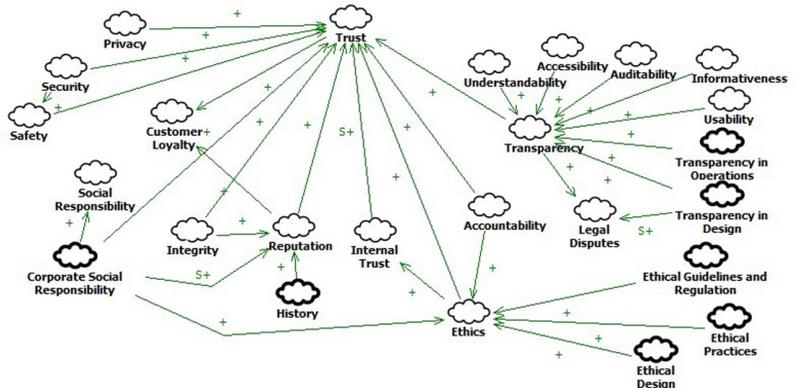

Fig. 1. High-Level Catalogue

(E)Charitable giving can either help or harm the reputation and Trust in an organization, depending on the apparent motivations behind such giving. Ethically-motivated giving helps reputation and Trust, while profit-motivated giving, on the other hand, has the opposite effect [25].

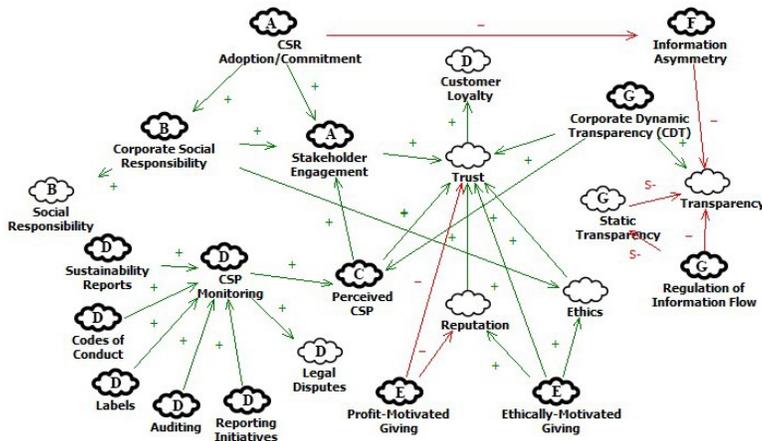

Fig. 2. Detailed Catalogue – Corporate Social Responsibility

(F)Cui et al., in their investigation into the effects of CSR adoption on Information Asymmetry, argue that increased CSR disclosure will reduce Information Asymmetry, which improves Transparency [26]. In Figure 2 this is mapped as a negative contribution to Information Asymmetry, and since Information Asymmetry contributes negatively to Transparency, diminishing information asymmetry increases Transparency. This extends to both financial and non-financial information, and any additional information provided from the organization to stakeholders reduces informational asymmetries between an organization and its stakeholders [26].

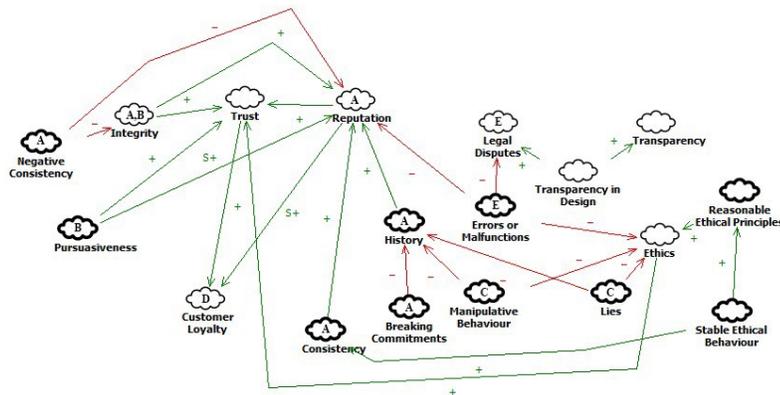

Fig. 3. Detailed Catalogue – Reputation

(G)Corporate Dynamic Transparency (CDT) is another specialized form of CSR. It specifically refers to the use of Information Communication Technology (ICT) to facilitate two-way exchange between corporations and stakeholders [27]. Vaccaro and Madsen demonstrated that this dynamic Transparency is both more desirable and more effective than the more common "static transparency", that is Transparency where a company's information disclosure is one-way [27]. This information disclosure via static Transparency is usually in response to government regulation and fosters less Trust than dynamic Transparency [27].

C. Reputation

(A)Reputation, in the context of our discussion on Trust, is comprised of three aspects: history, consistency, and errors or failures. History refers to the abstract sum-total of previous actions taken by the company that may positively or negatively influence its relationship with stakeholders. The history of interactions that a company has with society and its stakeholders helps Trust, assuming that history is positive [28]. A company's history also affects its reputation since previous customers can communicate this reputation to other customers, even those the organization may have no previous history with. Furthermore, the company's consistency in question may positively affect its reputation, and vice-versa for negative actions [28]. The reputation of a company is closely linked to stakeholders' Trust in a company [28]. A positive reputation helps Trust, while a negative reputation hurts Trust [28]. Breaking commitments or promises hurts both the history and, subsequently, the reputation of a company, which in term hurts Trust [28]. Negative consistency, which is consistently performing actions that would negatively affect a company's reputation, hurts a company's integrity and results in lost stakeholders' Trust [28].

(B)Ayaburi and Treku, in their study on loss of Trust in Facebook and subsequent attempts to repair that Trust, found that the effectiveness of trust repair is heavily dependent on the persuasiveness of the apology, as well as the integrity of the organization issuing the apology [3].

(C)The stability of ethical behaviour and effective patterns of ethical behaviour also help foster Trust [29]. The key is stability over time. A history of lies counteracts occasional public statements and acts of Transparency [29], [48]. History of untrustworthiness, unethicalness, lies, or manipulative behaviour will severely harm Trust [29], [48].

(D)Customer loyalty is a softgoal that is closely intertwined with the practices of CSR. Loyalty itself can be split into purchase loyalty and attitudinal loyalty. It is Trust, specifically Trust in a brand, and the brand effect that builds customer loyalty [10]. The link between loyalty and reputation needs further investigation, but Park et al. mention that reputation has some influence on loyalty and how customers make purchase decisions [11].

(E)An inevitability of software is the occurrence of errors or malfunctions. In these cases, Transparency in software can help determine responsibilities and liabilities, as well as civil and criminal implications [6], [30]. Furthermore, especially when dealing with safety-critical software that can profoundly impact human lives or even have the potential to end human lives, the need for a company to minimize errors, malfunctions and maintain high-quality standards becomes an ethical matter

[6]. The company's responsibility is to adhere to the highest quality standards [6], [30].

*D. Internal Trust*

(A)Trust as a concept, as we have been discussing so far, refers to the external Trust between a company and its stakeholders, often consumers. Internal Trust refers to the Trust between a company and its employees, regardless of whether those employees are stakeholders or not. Contrary to the softgoal of external Trust, which is our main focus, internal Trust is a subgoal that can affect external trust [28]. Internal Trust, through natural social interaction, benefits external Trust to some degree. Managers can be trained to enhance their trustworthiness and, thus, internal Trust through trust-training [28]. Trust-training exposes managers to the phenomenon of Trust and helps them develop some understanding of trust dynamics, including what trust-formation entails [28]. Ethical behaviour plays a significant role in this process, and the ethical facilitators of openness, integrity, benevolence, and history of interactions play a role in fostering internal Trust [28].

(B)Managerial competency also builds internal Trust. This refers not to technical expertise but rather the managers' competency to behave ethically, manage people and affairs in a morally sensitive way, and cultivate Trust within their subordinates [28].

(C)Company downsizing can harm internal Trust, as it can lead to fear within employees of losing their jobs [28]. Internal communication between management and employees and management's adequacy in informing employees of information that the employees consider important is an important factor in cultivating trust [28]. Openness and honesty play a large part in this communication [28]. Too much openness resulting in either a breach of confidentiality or in burdening staff with one's personal life can erode Trust. Furthermore, the poor history between the employees and the company, including broken commitments, will also erode trust [28].

*E. Ethics*

Three categories pertaining to Ethics were elicited during open coding. These are Ethically Motivated Practices, Ethically Motivated Guidelines and Regulation, and Ethical Software Design. Although these three categories fall under the umbrella of Ethics, we found that they were distinct enough concepts to warrant their own categories during open coding. However, due to the quantity and complexity of interrelationships between the operationalizations that compose these sub-categories, we have chosen to represent all three sub-categories and their decomposed operationalizations on one SIG rather than three separate SIGs to facilitate understandability, reduce redundancy, and increase clarity.

(A)Lin postulates that "reasonable ethical principles", such as those principles or actions aiming to save the greatest number of lives are a reasonable approach to Ethics [6]. Exhibiting social morality in this way is generally considered to be a positive approach to Ethics [49].

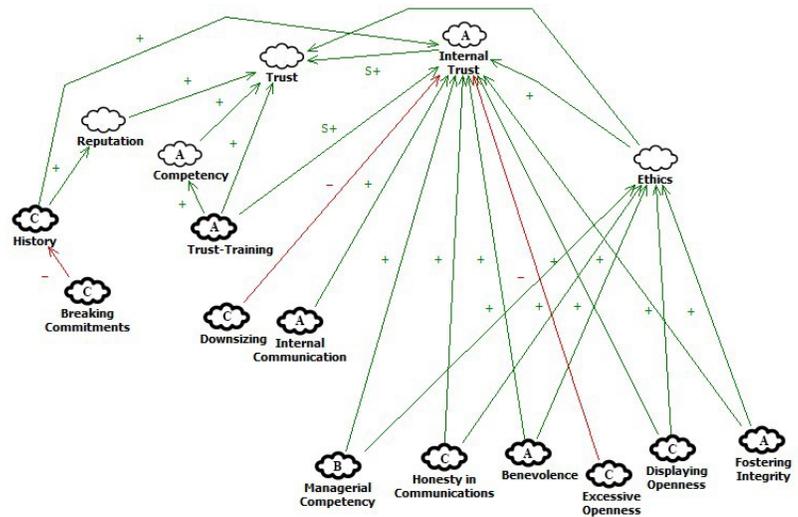

Fig. 4. Detailed Catalogue – Internal Trust

(B)The state of being ethically sound relies on having and exhibiting reasonable ethical principles [7]. Societal morals rather than individual Ethics drives these principles, and the communication of such ethical principles to stakeholders is vital to being ethically sound [7], [28]. Bews and Rossouw say that clear ethical guidelines help both Transparency and Trust but notes that it is important to understand this does not "make" Ethics since good ethical guidelines are detrimental to trust when not followed correctly [28]. Furthermore, not following guidelines leads to a negative reputation and breaks Trust [28]. Bachmann et al. postulates that in cases of lost Trust, trust repair requires formal rules, controls, and regulation to constrain untrustworthy behaviour to prevent further future trust violations [2].

(C)Lin takes this further and says communication is a restriction on ethical principles. Ethical principles should be clearly communicated and acted upon [6]. Communication itself is vital, and Lin mentions a reasonable principle, such as aiming to save the greatest number of lives can be justified as a feature and not a bug [6]. Bachmann et al. reinforces this by stating that organizations and institutions, in their work to investigate methods to repair Trust, suggest that trustworthiness can be signaled to stakeholders by developing and communicating a strong ethical culture [2]. Kang and Hustvedt found that socially responsible companies that give back to their local community enjoy increased consumer trust and increased positive word-of-mouth [31].

(D)Ethical facilitators of trustworthiness include benevolence, integrity, competency, openness, concern, reliability, and loyalty [28], [32]. Bews and Rossouw hypothesize that displaying openness, fostering integrity, and exhibiting benevolence are all primary ethical facilitators of fostering trust [28]. Fostering integrity can be further divided into three primary operationalizations that result in increased integrity [28]. These are consistency, at least positive consistency, reliability, and fairness [28]. Consistency has already been discussed in a previous section. Having a high level of reliability is often associated with trustworthiness [28]. Fairness refers to the act of giving equal moral considering to the interests of others, and a commitment to fairness or the exhibition of fairness increases Trust [28].

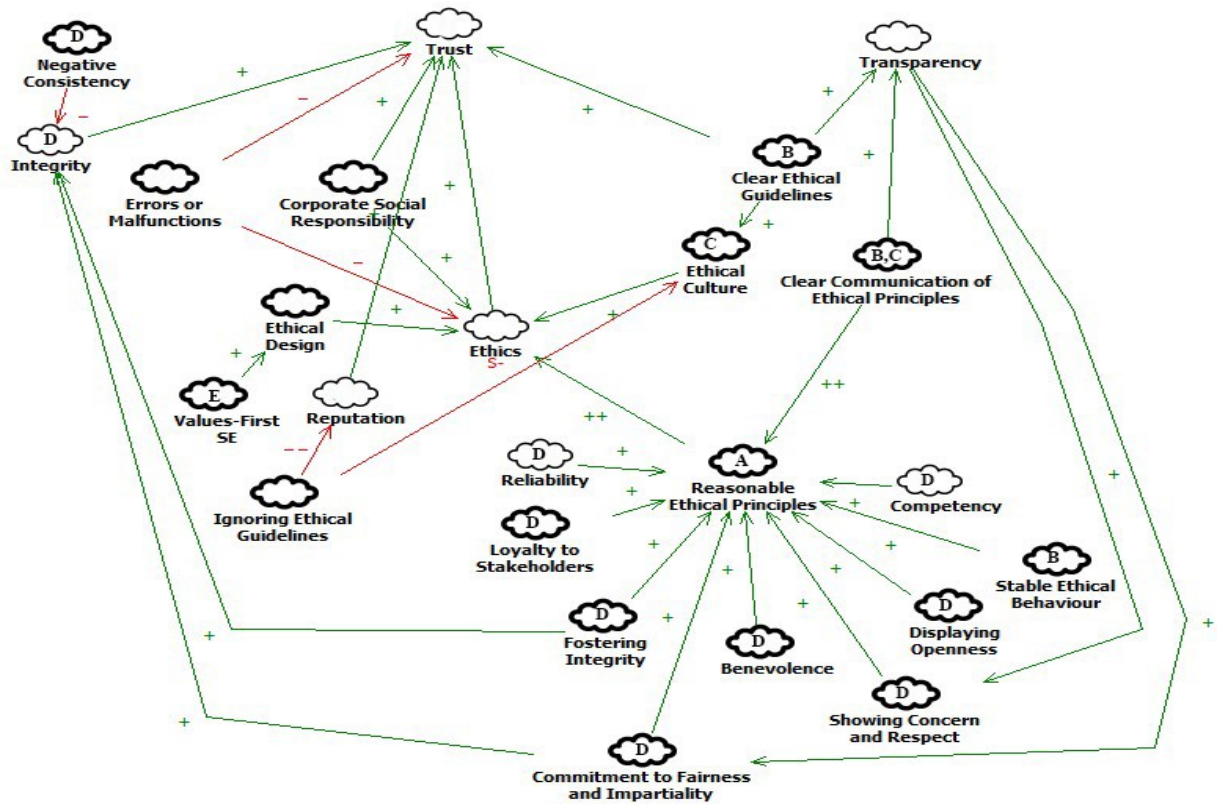

Fig. 5. Detailed Catalogue – Ethics

(E)Values-First software engineering is a software engineering practice that focuses on embedding values research principles into the software engineering decision-making process while also extracting lessons learned from practice [34]. Since Trust is one of these driving values, values-first software engineering intrinsically emphasizes helping foster Trust [34].

### F. Transparency in Operations

Two categories pertaining to Transparency were elicited during open coding: Transparency in Operations and Transparency in Design. We have chosen to represent both of these sub-categories and their decomposed operationalizations on one SIG to facilitate understandability of their relationships.

Our literature review found that Transparency's subgoal has a helpful impact on Trust in the general case. This is in line with the preliminary SIG proposed by Cysneiros and Leite [9]. Transparency takes many forms, and different forms of Transparency in different conditions may have varying effects. Depending on the author and the specific circumstances and context that the author is investigating Transparency in, we found that Transparency can also harm Trust. In this SIG, we present both helpful and hurtful operationalizations regarding the effects of Transparency on Trust and show the operationalizations that lead to either positive or negative effects on both Transparency and Trust.

(A)Leite and Cappelli created a catalogue of Transparency with a set of possible solutions to build transparent software, as well as the role of the NFRs of understandability, auditability, accessibility, usability, and informativeness on transparency [35]. We include these NFRs in our detailed view of Transparency related to other operationalizations in other sections, as can be seen on those SIGs.

(B)Transparency itself is unlikely to produce lasting Trust without certain protections. Corporations need not explicitly express this Transparency regarding stakeholders' rights, but they must exhibit themselves to care about those rights [33].

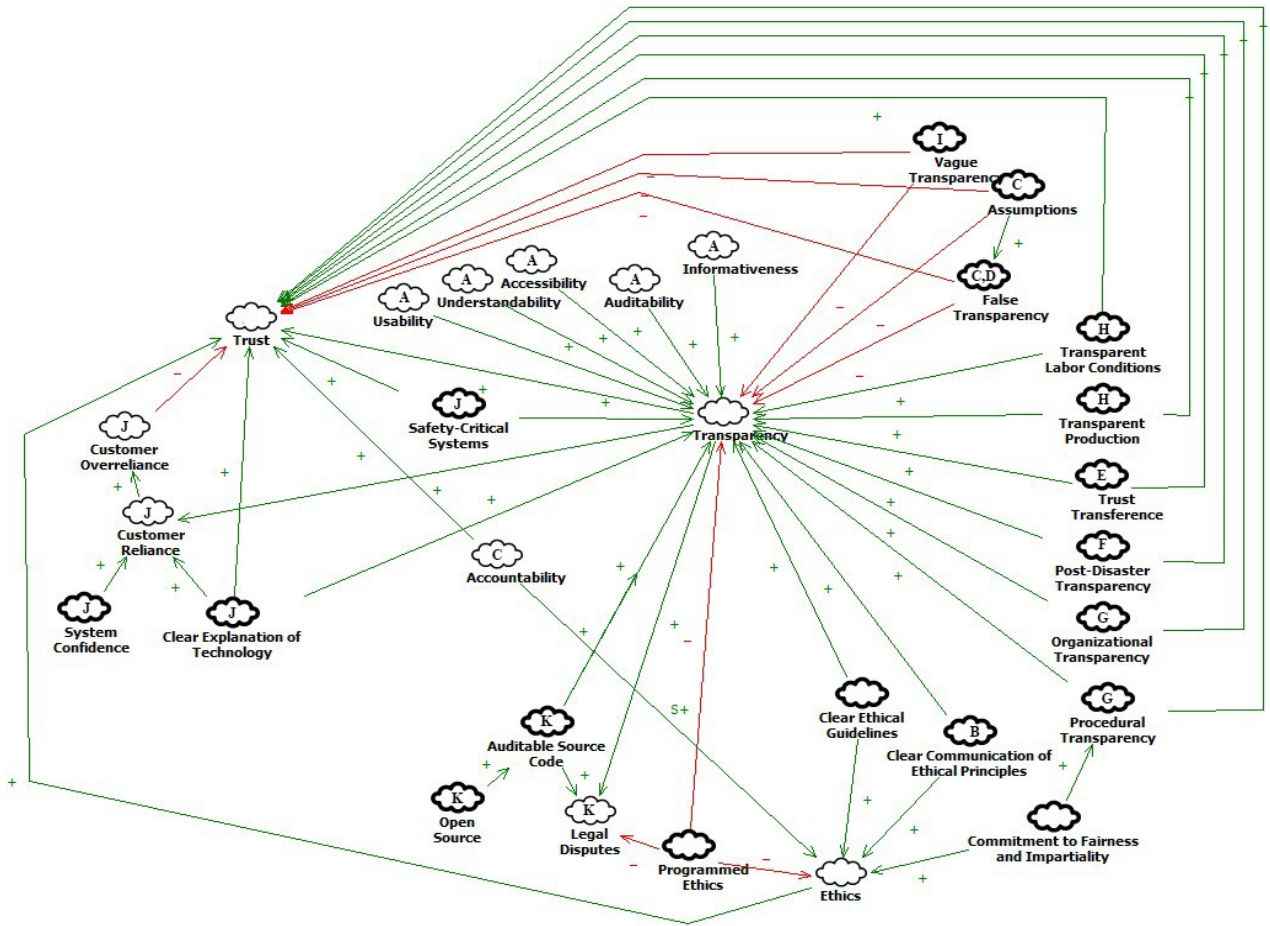

Fig. 6. Detailed Catalogue – Transparency

[C]Bachmann et al., in their work on trust repair, suggests that there are two aspects to Transparency, that of information sharing and that of accountability [2]. The Global Reporting Initiative provides guidelines that organizations can follow to become more accountable and transparent. Child and Rodrigues suggest that organizations and institutions that embrace this idea and comply with principles of accountability, Transparency, and disclosure receive increased or restored Trust [36]. Bachmann et al. also argues that making organizational information transparent results in a firmer basis for stakeholders to evaluate an organization's trustworthiness [2]. The disclosed information may reveal organizational competence, integrity, and benevolence, or even reveal the opposite [2]. As a trust repair mechanism, Transparency requires that stakeholders trust that the organization is honest, comprehensive, and balanced in its transparent reporting [2]. An organization that discloses only positive information without disclosing negative information would be inherently untrustworthy (i.e., false Transparency) [2]. Furthermore, when assumptions are made, particularly false assumptions, about the information that is not transparent, Transparency loses its value, and false Transparency becomes an issue [2].

[D]Santana and Wood argue that the appearance of Transparency is not enough to foster Trust if the Transparency is not genuine [37]. Santana and Wood use the case study of Wikipedia, which they argue has the appearance but not the reality of responsible and transparent information production. Wikipedia, Santana and Wood argue, has highly transparent writing and editing processes but severely lacks Transparency regarding the contributors' identity, that is, the writers, editors, and administrators [37]. This appearance of Transparency is not real Transparency and ultimately hurts legitimacy and validity [37].

[E]Trust can be transferred from a credible, trusted actor or organization to a discredited, scandalized, or untrustworthy actor or organization [38]. Trust transferability is the central mechanism of trust creation (or re-creation) in the process of trust repair [38]. One of the primary trust transference methods is exhibited in trust breaches by auditing firms [38]. Through a public inquiry process, Trust and legitimacy were transferred from the government's parliamentary inquiry leaders, since they were impartial, to the auditing firms after a significant structural overhaul of the audit system [38]. This public inquiry process is a significant exercise in Transparency, and

thus Trust transference helps both Transparency and Trust [2], [38].

**(F)**In situations where Trust is already broken or lost, where there is already a negative history of trustworthy actions, what Bachmann et al. calls post-disaster Transparency can help mend Trust [2]. Furthermore, Kim et al., while examining whether CSR mitigates the risk of stock price crashes, found that if socially responsible firms commit themselves to a high standard of Transparency, mainly by not withholding bad news, they have a lower crash risk from stakeholder dissatisfaction and lack of faith [39].

**(G)**Organizations that promote procedural transparency experience increased Trust [28]. Transparency in how an organization handles internal procedures such as promotions, bonuses, disciplinary actions, and layoffs leads to increased internal Trust as long as they are handled fairly and consistently [28]. In their 2012 dissertation focusing on microfinance firms, Augustine found both qualitative and quantitative evidence that organizational Transparency, a vital aspect of good corporate governance, results in positive implications in firm performance [40]. Augustine's evidence points to organizational Transparency leading directly to stakeholder trust in the microfinance firm since Transparency delivers information, procedures, and services that facilitate the relationship between microfinance firms and their stakeholders [40].

**(H)**Transparent production refers to the clear communication of how companies create or acquire their products. We consider this also to include the operations of service-oriented companies regarding Transparency in how they provide their service and platform-oriented companies regarding Transparency in how the platform operates. Kang and Hustvedt found that transparent companies in their production and labor conditions enjoyed increased Trust from consumers and increased positive word-of-mouth regarding the company and its products [31]. In an example of Transparency that fosters Trust and privacy, Anthonysamy et al. studied social network providers and experience enhanced user trust when they demonstrate they have taken adequate measures in their platform operations to protect users' personal information [41].

**(I)**Transparency must have the right level of specificity to enhance trust [29], [42]. Eslami et al. found that Transparency about the existence of a newsfeed algorithm required a certain level of specificity and resulted in positive reactions in test subjects [42]. Felzmann suggests that transparent explanations that are too vague or too specific incites feelings of unease and distrust and decrease stakeholders' Trust and satisfaction [29].

### G. Transparency in Design

**(J)**In their work on Trust and reliance in clinical decision support systems, Bussone et al. found that a clear or fuller explanation of the facts and/or inner workings of technology helps with both the Trust and reliance on the system [43]. Furthermore, in systems that can have a confidence level in its own decisions, such as clinical decision support systems, a higher confidence level results in a slightly higher level of Trust in the system [43]. The level of reliance that stakeholders' have on the system also increases their Trust in the system, possibly out of necessity. As long as some or all of these criteria are satisfied, clinical decision support systems enjoy client trust. However, overreliance, or misplaced Trust, and the reveal thereof, can have drastic negative consequences on trust [43]. If it is revealed that the Trust that consumers place in technology is underserving, then the result is a catastrophic loss of Trust [43]. Furthermore, Bowen warns that decisions in the software-engineering design process are often based on economic rather than safety considerations [7]. For safety-critical systems, where the importance of safety is paramount, stakeholders' inherent Trust in the product may be unfounded [7].

**(K)**In 2016, Walker stated that our use of technology as we progress through the digital age is increasing faster than our understanding of the issues involved [44]. Walker urges that surrendering our personal information to technology is an increasing problem in society. However, it can be mediated if policymakers implement effective regulation to increase Transparency and attention and reduce uncertainty regarding how the technology handles our personal information [44]. In their work with Implantable Medical Devices (IMDs), Sandler found that source-code that is auditable helps with stakeholders' Trust and the ability to handle legal disputes [8]. In fact, they strongly argue for the source-code of IMDs to be open source and publicly auditable [8]. Sandler argues that legal regulation that demands open-source access to IMDs, or at the very least, federal access to source-code, will improve auditability, help prevent catastrophic failure, and drive greater Transparency [8].

## IV. THREATS TO VALIDITY

The critical aspect of our analysis was the identification and synthesis of complex concepts from existing literature. It should be acknowledged that this process is inseparably linked to the researchers, so the question arises as to whether the SIGs are a valid representation of the analyzed concepts or if they contain some reflection of the researchers' own views. This is a threat to construct validity, albeit one that is inherent to qualitative research. Our usage of Wolfswinkel et al.'s grounded theory approach to literature review was vital in our efforts to minimize any preconceptions we may have had as researchers and let the data speak for itself. Traditional grounded theory proposes that researchers enter the process with no preconceptions with regards to the relevance of particular information. More modern views on grounded theory however, propose that preconceptions are difficult or nearly impossible to avoid completely, and rather it is the attempt to minimize the effect of these preconceptions that is vital. To this end we attempted to represent all viewpoints found in the analyzed literature. Furthermore, although only one researcher participated in the matrix-creation steps of coding, we attempted to minimize this threat by following a process whereby one researcher translated the coding results into SIGs while two other researchers analyzed the results to create the final SIG.

A threat to external validity exists regarding the generalizability of our findings. Due to the lack of existing literature focusing directly on software-related social Trust and how this affects business performance, we have developed our catalogue using literature primarily from business and social sciences. It is assumed that a company's identity is intricately linked with its products, and thus trust in a company leads to trust in its products. Based on this assumption, it is reasonable to assert that this research presents a solid basis for further investigating the effects of Transparency and Ethics upon software trust. However, we have only superficially looked at Ethics through the viewpoint of fairness. We intend to address that in future work.

## V. Discussion

Following the 2014 Cambridge Analytics scandal, Facebook users' personal information (PI) and opinions were harvested and used without authorization. Facebook experienced a loss of consumer trust [3]. This was followed by a decrease in its platform's use and heightened concern for privacy from its users [3]. To regain consumer trust, Facebook offered a public apology and invested in an extensive marketing campaign in 2018 [3]. What if, in that case, catalogues such as we described were used? We argue that if used, the catalogs could have helped Facebook avoid a loss of consumer trust.

This paper presented a set of possible operationalizations (solutions) that could have been implemented before and after the scandal in 2014. As detailed by the Ethics catalogue, Facebook could have exhibited reasonable ethical principles by not participating in PI's unauthorized harvesting. Clear communication of these ethical principles would have increased Transparency and Trust. If Facebook was resigned to participating in PI's harvesting, they could have followed a transparent harvesting method as detailed by the Transparency catalogue. Facebook could have exhibited procedural Transparency and transparent production to protect Trust in the company. They could have avoided any cases of vague Transparency or false Transparency to avoid any further loss of Trust. Lastly, as exhibited by the Reputation catalogue, Facebook could have increased Trust with their apology's persuasiveness and the consistency of which they adhere to PI protection.

Although such an alternate timeline is merely conjecture, we believe it is a step in the right direction towards the advancement of SRS. SIG catalogues have been found to be effective in aiding requirements engineers in satisfying targeted NFRs [50], [51] as well as uncovering goals that may have not been previously elicited [52]. Further work should investigate systematic methods of utilizing SIG catalogues, as well as the validation of the efficacy of specific catalogues such as the one presented in this paper.

Furthermore, future work should also investigate the pros and cons of ethical dilemmas that have no morally clear solution and how different courses of action affect Trust and legal disputes. This is challenging since many of these ethical dilemmas, such as with autonomous vehicles, have no current legislation or case law to support one side or another, and there are still debates under contention.

Nevertheless, this work already exposes possible solutions and the effects each of them may cause, bringing to light a line of reasoning that could orient our decisions towards creating trustworthy software. Our catalogue shows how social responsibility can be extrapolated to software, software companies, and companies that create software-embedded products. If companies can better understand how Ethics, Transparency, and other NFRs drive consumer trust in them, then by extension, their products will be trustworthy as well. We believe this catalogue will aid software engineers, and by extension, stakeholders, in handling costly litigation, attracting new clients and retaining current ones.

## VI. Conclusion

This work shows a first approach to bring new emerging concepts in society such as Transparency and Ethics into the requirements engineering process. It relates these aspects aiming at achieving Trust, which is a key principle to CSR. We extrapolate the CSR, inspired by Cysneiros and Leite's work [9], in a way to support the concept of software with social responsibility. For that, we developed an initial NFR catalogue into the effects of Transparency and Ethics on Trust. We found that ethical behaviors are considered to positively affect Trust, at least from the perspective of society's morals.

Unlike Ethics, where only morally ambiguous dilemmas have drastically different viewpoints between authors, we found that Transparency's general effects on Trust are much less clear-cut. Our research indicates that increases in software or organizational Transparency generally help promote Trust, and Transparency only harms Trust under special circumstances [3].

Although far from exhaustive, our catalogue presents an extensive set of options that could be used in different situations by software engineers. The options presented here do not seek to highlight one specific or correct option to operationalize Trust, but rather to present various alternatives in one knowledge base. In future work, we plan to integrate existing NFR catalogues further. This means continue to elicit knowledge from the literature to enrich the catalogs, like exploring Ethics from inside out (for instance, as fairness) and exposing the results to practitioners and getting feedback from the trenches likewise the work of Sadi and Yu [53].

REFERENCES

[1] O. Zinovatna and L. M. Cysneiros, "Reusing knowledge on delivering privacy and transparency together," *Requir. Patterns (RePa), 2015 IEEE Fifth Int. Work.*, pp. 17–24, 2015, doi: 10.1109/RePa.2015.7407733.

[2] R. Bachmann, N. Gillespie, and R. Priem, "Repairing Trust in Organizations and Institutions: Toward a Conceptual Framework," *Organ. Stud.*, vol. 36, no. 9, pp. 1123–1142, Sep. 2015, doi: 10.1177/0170840615599334.

[3] E. W. Ayaburi and D. N. Treku, "Effect of penitence on social media trust and privacy concerns: The case of Facebook," *Int. J. Inf. Manage.*, vol. 50, pp. 171–181, 2020, doi: https://doi.org/10.1016/j.ijinfomgt.2019.05.014.

[4] M. J. Hanna and S. C. Kimmel, "Current US Federal Policy

Framework for Self-Driving Vehicles: Opportunities and Challenges," *Computer (Long. Beach. Calif).*, vol. 50, no. 12, pp. 32–40, 2017, doi: 10.1109/MC.2017.4451211.

[5] D. Morris, "Mercedes' Self-Driving Cars Will Save Passengers, Not Bystanders," *Fortune.com*, Oct. 15, 2016. https://fortune.com/2016/10/15/mercedes-self-driving-car-ethics/ (accessed Aug. 13, 2020).

[6] P. Lin, "Why Ethics Matters for Autonomous Cars," in *Autonomes Fahren*, Berlin, Heidelberg: Springer Berlin Heidelberg, 2015, pp. 69–85.

[7] J. Bowen and Jonathan, "The ethics of safety-critical systems," *Commun. ACM*, vol. 43, no. 4, pp. 91–97, Apr. 2000, doi: 10.1145/332051.332078.

[8] K. Sandler, L. Ohrstrom, L. Moy, and R. Mcvay, "Killed by Code: Software Transparency in Implantable Medical Devices," *Broadway*, vol. 5882, pp. 1–212, 1995, [Online]. Available: www.softwarefreedom.org.

[9] L. M. Cysneiros and J. C. S. do Prado Leite, "Non-functional requirements orienting the development of socially responsible software," in *Lecture Notes in Business Information Processing*, 2020, vol. 387 LNBIP, pp. 335–342, doi: 10.1007/978-3-030-49418-6_23.

[10] A. Chaudhuri and M. B. Holbrook, "The Chain of Effects from Brand Trust and Brand Affect to Brand Performance: The Role of Brand Loyalty," *J. Mark.*, vol. 65, no. 2, pp. 81–93, Apr. 2001, doi: 10.1509/jmkg.65.2.81.18255.

[11] E. Park, K. J. Kim, and S. J. Kwon, "Corporate social responsibility as a determinant of consumer loyalty: An examination of ethical standard, satisfaction, and trust," *J. Bus. Res.*, vol. 76, pp. 8–13, Jul. 2017, doi: 10.1016/J.JBUSRES.2017.02.017.

[12] L. Chung, B. A. Nixon, E. Yu, and J. Mylopoulos, *Non-Functional Requirements in Software Engineering*. Springer US, 2000.

[13] F. B. Aydemir and F. Dalpiaz, "A Roadmap for Ethics-Aware Software Engineering," in *2018 IEEE/ACM International Workshop on Software Fairness (FairWare)*, 2018, pp. 15–21, doi: 10.23919/FAIRWARE.2018.8452915.

[14] A. Strauss and J. Corbin, *Basics of Qualitative Research: Techniques and Grounded Theory Procedures for Developing Grounded Theory*. 1998.

[15] O. Badreddin, "Thematic Review and Analysis of Grounded Theory Application in Software Engineering," *Adv. Softw. Eng.*, vol. 2013, pp. 1–9, 2013, doi: 10.1155/2013/468021.

[16] D. Würfel, R. Lutz, and S. Diehl, "Grounded requirements engineering: An approach to use case driven requirements engineering," *J. Syst. Softw.*, vol. 117, pp. 645–657, 2016, doi: https://doi.org/10.1016/j.jss.2015.10.024.

[17] J. F. Wolfswinkel, E. Furtmueller, and C. P. M. Wilderom, "Using grounded theory as a method for rigorously reviewing literature," *European Journal of Information Systems*. 2013, doi: 10.1057/ejis.2011.51.

[18] J. Pivatelli and J. C. S. do Prado Leite, "The Clash between Requirements Volatility and Software Contracts," in *Proceedings of the 31st Brazilian Symposium on Software Engineering*, 2017, pp. 144–153, doi: 10.1145/3131151.3131159.

[19] H. Jo and M. A. Harjoto, "Corporate Governance and Firm Value: The Impact of Corporate Social Responsibility," *J. Bus. Ethics*, vol. 103, no. 3, pp. 351–383, 2011, doi: 10.1007/s10551-011-0869-y.

[20] H. Jo and M. A. Harjoto, "The Causal Effect of Corporate Governance on Corporate Social Responsibility," *J. Bus. Ethics*, vol. 106, no. 1, pp. 53–72, 2012, doi: 10.1007/s10551-011-1052-1.

[21] B. E. Joyner and D. Payne, "Evolution and implementation: A study of values, business ethics and corporate social responsibility," *J. Bus. Ethics*, vol. 41, no. 4, pp. 297–311, 2002, doi: 10.1023/A:1021237420663.

[22] J. M. Rose, "Corporate directors and social responsibility: Ethics versus shareholder value," *J. Bus. Ethics*, vol. 73, no. 3, pp. 319–331, Jul. 2007, doi: 10.1007/s10551-006-9209-z.

[23] D. J. Wood, "Corporate Social Performance Revisited," *Acad. Manag. Rev.*, vol. 16, no. 4, pp. 691–718, Oct. 1991, doi: 10.5465/amr.1991.4279616.

[24] T. Talaulicar, "Barriers Against Globalizing Corporate Ethics: An Analysis of Legal Disputes on Implementing U.S. Codes of Ethics in Germany," *J. Bus. Ethics*, vol. 84, no. 3, p. 349, 2009, doi: 10.1007/s10551-009-0199-5.

[25] P. A. Vlachos, A. Tsamakos, A. P. Vrechopoulos, and P. K. Avramidis, "Corporate social responsibility: Attributions, loyalty, and the mediating role of trust," *J. Acad. Mark. Sci.*, vol. 37, no. 2, pp. 170–180, Jun. 2009, doi: 10.1007/s11747-008-0117-x.

[26] J. Cui, H. Jo, and H. Na, "Does Corporate Social Responsibility Affect Information Asymmetry?," *J. Bus. Ethics*, vol. 148, no. 3, pp. 549–572, Mar. 2018, doi: 10.1007/s10551-015-3003-8.

[27] A. Vaccaro and P. Madsen, "Corporate dynamic transparency: The new ICT-driven ethics?," *Ethics Inf. Technol.*, vol. 11, no. 2, pp. 113–122, 2009, doi: 10.1007/s10676-009-9190-1.

[28] N. F. Bews and G. J. Rossouw, "A Role for Business Ethics in Facilitating Trustworthiness," *J. Bus. Ethics*, vol. 39, no. 4, pp. 377–390, 2002, doi: 10.1023/A:1019700704414.

[29] H. Felzmann, E. F. Villaronga, C. Lutz, and A. Tamò-Larrieux, "Transparency you can trust: Transparency requirements for artificial intelligence between legal norms and contextual concerns," *Big Data Soc.*, vol. 6, no. 1, pp. 1–14, 2019, doi: 10.1177/2053951719860542.

[30] A. Hevelke and J. Nida-Rümelin, "Responsibility for Crashes of Autonomous Vehicles: An Ethical Analysis," *Sci. Eng. Ethics*, vol. 21, no. 3, pp. 619–630, 2015, doi: 10.1007/s11948-014-9565-5.

[31] J. Kang and G. Hustvedt, "Building Trust Between Consumers and Corporations: The Role of Consumer Perceptions of Transparency and Social Responsibility," *J. Bus. Ethics*, vol. 125, no. 2, pp. 253–265, Dec. 2014, doi: 10.1007/s10551-013-1916-7.

[32] F. D. Schoorman, R. C. Mayer, and J. H. Davis, "An Integrative Model of Organizational Trust: Past, Present, and Future," *Acad. Manag. Rev.*, vol. 32, no. 2, pp. 344–354, Apr. 2007, doi: 10.5465/amr.2007.24348410.

[33] J. Elia, "Transparency rights, technology, and trust," *Ethics Inf. Technol.*, vol. 11, no. 2, pp. 145–153, 2009, doi: 10.1007/s10676-009-9192-z.

[34] M. Angela Ferrario, W. Simm, S. Forshaw, A. Gradinar, M. Tavares Smith, and I. Smith, "Values-First SE: Research Principles in Practice," *IEEE/ACM 38th Int. Conf. Softw. Eng. Companion*, pp. 553–562, 2016, doi: 10.1145/2889160.2889219.

[35] J. Leite and C. Cappelli, *C.S.and Software Transparency," Business & Information Systems Engineering: Vol. Iss. 3, . Available at: http://aisel.aisnet.org/bise/vol2/iss3/3*, vol. 2. 2010.

[36] J. Child and S. B. Rodrigues, "Repairing the Breach of Trust in Corporate Governance," *Corp. Gov. An Int. Rev.*, vol. 12, no. 2, pp. 143–152, Apr. 2004, doi: 10.1111/j.1467-

8683.2004.00353.x.
[37] A. Santana and D. J. Wood, "Transparency and social responsibility issues for Wikipedia," *Ethics Inf. Technol.*, vol. 11, no. 2, pp. 133–144, 2009, doi: 10.1007/s10676-009-9193-y.
[38] F. Mueller, C. Carter, and A. Whittle, "Can Audit (Still) be Trusted?," *Organ. Stud.*, vol. 36, no. 9, pp. 1171–1203, Jun. 2015, doi: 10.1177/0170840615585336.
[39] Y. Kim, H. Li, and S. Li, "Corporate social responsibility and stock price crash risk," *J. Bank. Financ.*, vol. 43, pp. 1–13, 2014, doi: https://doi.org/10.1016/j.jbankfin.2014.02.013.
[40] D. Augustine, "Good Practice in Corporate Governance," *Bus. Soc.*, vol. 51, no. 4, pp. 659–676, Dec. 2012, doi: 10.1177/0007650312448623.
[41] P. Anthonysamy, P. Greenwood, and A. Rashid, "Social networking privacy: Understanding the disconnect from policy to controls," *Computer (Long. Beach. Calif).*, vol. 46, no. 6, pp. 60–67, 2013.
[42] M. Eslami *et al.*, "'I Always Assumed That I Wasn't Really That Close to [Her]': Reasoning about Invisible Algorithms in News Feeds," in *Proceedings of the 33rd Annual ACM Conference on Human Factors in Computing Systems*, 2015, pp. 153–162, doi: 10.1145/2702123.2702556.
[43] A. Bussone, S. Stumpf, and D. O'Sullivan, "The role of explanations on trust and reliance in clinical decision support systems," in *Proceedings - 2015 IEEE International Conference on Healthcare Informatics, ICHI 2015*, Dec. 2015, pp. 160–169, doi: 10.1109/ICHI.2015.26.
[44] K. L. Walker, "Surrendering information through the looking glass: Transparency, trust, and protection," *J. Public Policy Mark.*, vol. 35, no. 1, pp. 144–158, Apr. 2016, doi: 10.1509/jppm.15.020.
[45] E. Delgado-Ballester, J. L. Munuera-Aleman, and M. J. Yague-Guillen, "Development and validation of a brand trust scale," *Int. J. Mark. Res.*, vol. 45, no. 1, pp. 35–56, Mar. 2003, [Online]. Available: http://go.galegroup.com/ps/anonymous?id=GALE%7CA97176865&sid=googleScholar&v=2.1&it=r&linkaccess=abs&issn=14707853&p=AONE&sw=w.
[46] S. Pivato, N. Misani, and A. Tencati, "The impact of corporate social responsibility on consumer trust: the case of organic food," *Bus. Ethics A Eur. Rev.*, vol. 17, no. 1, pp. 3–12, Dec. 2007, doi: 10.1111/j.1467-8608.2008.00515.x.
[47] J. Choi and H. Wang, "Stakeholder relations and the persistence of corporate financial performance," *Strateg. Manag. J.*, vol. 30, no. 8, pp. 895–907, Aug. 2009, doi: 10.1002/smj.759.
[48] S. Zuboff, *The Age of Surveillance Capitalism: The Fight for the Future at the New Frontier of Power*. New York: PublicAffairs, 2019.
[49] S. Hampshire, *Morality and Conflict*. Harvard University Press, 1984.
[50] L. M. Cysneiros, "Evaluating the Effectiveness of Using Catalogues to Elicit Non-FunctionalRequirements.," in *WER*, 2007, pp. 107–115.
[51] M. de Gramatica, K. Labunets, F. Massacci, F. Paci, and A. Tedeschi, "The Role of Catalogues of Threats and Security Controls in Security Risk Assessment: An Empirical Study with ATM Professionals," Springer International Publishing, 2015, pp. 98–114.
[52] E. Cardoso, J. P. A. Almeida, R. S. S. Guizzardi, and G. Guizzardi, "A Method for Eliciting Goals for Business Process Models based on Non-Functional Requirements Catalogues," *Int. J. Inf. Syst. Model. Des.*, vol. 2, no. 2, pp. 1–18, Apr. 2011, doi: 10.4018/jismd.2011040101.
[53] M. H. Sadi and E. Yu, "RAPID: a knowledge-based assistant for designing web APIs," *Requir. Eng.*, vol. 1, p. 3, Feb. 2021, doi: 10.1007/s00766-020-00342-0.